\documentclass{iopart}%[nofootinbib,superscriptaddress]{iopart}
\usepackage{bbm}
\usepackage{mathrsfs}
\usepackage{mathptmx}
\usepackage{epsfig}
\usepackage{graphicx}
\usepackage{slashed}
\usepackage[compress]{cite}

\usepackage{color}

\def\comment#1{}

\begin{document}

\title[]{Surface tension of compressed, superheavy atoms}

\author{Jorge A. Rueda, Yuan-Bin Wu\footnote{send correspondence to wuyb@icranet.org}\footnote{Present address: Max-Planck-Institut f\"ur Kernphysik, Saupfercheckweg 1, D-69117 Heidelberg, Germany}, and She-Sheng Xue}
\address{Dipartimento di Fisica and ICRA, Sapienza Universit\`a
di Roma, Piazzale Aldo Moro 5, I-00185 Rome, Italy}%
\address{ICRANet, Piazza della Repubblica 10, I-65122 Pescara,
Italy}%

\begin{abstract}
Based on the relativistic mean field theory and the Thomas-Fermi approximation, 
we study the surface properties of compressed, superheavy atoms. 
By compressed, superheavy atom we mean an atom composed by 
a superheavy nuclear core (superheavy nucleus) with mass number 
of the order of $10^4$, and degenerate electrons that neutralize the system. 
Some electrons penetrate into the superheavy nuclear core and the rest surround it 
up to a distance that depends upon the compression level. Taking into account the
strong, weak, and electromagnetic interactions, we numerically study
the structure of compressed, superheavy atoms and calculate the
nuclear surface tension and Coulomb energy. We analyze the influence
of the electron component and the background matter on the nuclear
surface tension and Coulomb energy of compressed, superheavy atoms. 
We also compare and contrast these results in the case of
compressed, superheavy atoms with phenomenological results in
nuclear physics and the results of the core-crust interface of
neutron stars with global charge neutrality. Based on the numerical
results we study the instability against Bohr-Wheeler surface
deformations in the case of compressed, superheavy atoms. The
results in this article show the possibility of the existence of such 
compressed, superheavy atoms, and provide the evidence of strong effects of
the electromagnetic interaction and electrons on the structure of
compressed, superheavy atoms.

\end{abstract}

\pacs{21.10.-k, 05.30.Fk, 26.60.-c}
% corresponding to: (PACS 2010 version)
% Properties of nuclei; nuclear energy levels
% Fermion systems and electron gas (Quantum statistical mechanics)
% Nuclear matter aspects of neutron stars

%\maketitle

%\ioptwocol

%%%%%%%%%%%%%%%%%%%%%%%%%%%%%%%%%%%%%%%%%%%%%%%%%%%%%%%%%%%%%%%%%%%%%%
%%%%%%%%%%%%%%%%%%%%%%%%%%%%%%%%%%%%%%%%%%%%%%%%%%%%%%%%%%%%%%%%%%%%%%
\section{Introduction}
\label{sec:intro}
%%%%%%%%%%%%%%%%%%%%%%%%%%%%%%%%%%%%%%%%%%%%%%%%%%%%%%%%%%%%%%%%%%%%%%
%%%%%%%%%%%%%%%%%%%%%%%%%%%%%%%%%%%%%%%%%%%%%%%%%%%%%%%%%%%%%%%%%%%%%%

It has been shown recently that the Einstein-Maxwell-Thomas-Fermi
(EMTF) equations \cite{Rueda1} supersede the traditional
Tolman-Oppenheimer-Volkoff (TOV) \cite{Tolman1,Oppenheimer1}
equations used for the construction of neutron star equilibrium
configurations, when taking into account the strong, weak,
electromagnetic, and gravitational interactions. In contrast to the
imposing of the condition of local charge neutrality in the
traditional TOV approach, the condition of global charge neutrality
is applied in the EMTF approach, owing to the fact that the
traditional treatment imposing the condition of local charge
neutrality is not consistent with the field equations and
microphysical conditions of equilibrium for the system of neutrons,
protons, and electrons in $\beta$ equilibrium and obeying
relativistic quantum statistics \cite{RuedaPLB}.

In order to describe the strong interactions between nucleons, the
$\sigma$-$\omega$-$\rho$ nuclear model of relativistic mean field
theory (RMFT) \cite{Duerr1, Miller1, Walecka1,
Boguta1,Boguta2,Boguta3,Boguta4,Boguta5} is adopted in the EMTF
approach. This model contains Dirac nucleons together with a scalar
meson $\sigma$ and a vector meson $\omega$ as well as an isovector
meson $\rho$. The RMFT has gained great successes in giving a
quantitative description of nuclear properties
\cite{Serot1,Ring1,Bender1} and understanding the inhomogeneous
structures of low-density nuclear matter which can be realized in
supernovae cores or in neutron star crusts (see
e.g.~Refs.~\cite{Maruyama1, Oyamatsu2007, Avancini1, Okamoto1,
Grill1, Bao2014, Bao2015, Newton2009, Schuetrumpf2015, Sagert2016, Pais2016} about the nuclear pasta structures).

As shown in Ref.~\cite{Belvedere1}, the self-consistent solution of
the EMTF equations leads to a new structure of neutron stars, which
is significantly different from the neutron star structure obtained
from the TOV equations imposing local charge neutrality
\cite{Haensel1}. In this new structure of neutron stars, a
transition layer (interface) appears between the core and the crust
of the star, near the nuclear saturation density. There is a
discontinuity in the density at the core-crust transition in this
new structure of neutron stars. The core (bulk region) inside this
transition layer is a hadronic phase and the crust outside this
transition layer is composed of a nuclei lattice and relativistic
degenerate electrons and possibly neutrons at densities below the
nuclear saturation density and higher than the estimated
neutron-drip value $\sim 4.3\times 10^{11}$ g cm$^{-3}$ \cite{Baym1,
Baym2}. Inside the transition region, a very strong electric field
overwhelming the critical field $E_c=m^2_e c^3/(e \hbar)$ for vacuum
breakdown appears \cite{Belvedere1}, where $m_e$ is the electron
mass.

The surface properties of nuclear matter such as the surface tension
and the curvature energy play an important role in many situations
and phenomena such as the stability of nuclei, fragment
distributions in heavy-ion collisions, and phase transitions between
different phases of nuclear matter. The surface properties of
nuclear matter have been analyzed a lot in the past few decays for
the matter at the nuclear saturation density
\cite{Boguta2,Brack1,Sharma1,Eiff1,Eiff2,Eiff3,Centelles1,Estal2,Patra1,Danielewicz1},
as well as the matter at the supranuclear regime realized in the
interior of neutron stars \cite{Christiansen1,Alford1} for the phase
transition region and the pasta structures of the low-density
nuclear matter \cite{Maruyama1, Oyamatsu2007, Avancini1, Grill1}.

The surface properties of the core-crust interface of the new
neutron star structure obtained from the solution of the EMTF
equations has been studied in Ref.~\cite{RRWX2014} (see also
Ref.~\cite{Wu2014} for a brief description). We calculated in
Ref.~\cite{RRWX2014} the surface tension as well as the
electrostatic energy stored in this core-crust transition layer. We
analyzed the stability of these systems through the Bohr-Wheeler
fission mechanism \cite{Bohr1}. It was shown in Ref.~\cite{RRWX2014}
that the electromagnetic interaction and the presence of degenerate
electrons have evident effects on the surface properties of the
core-crust interface. In the analyses of
Refs.~\cite{Belvedere1, RRWX2014}, we employed the condition that
the electron density is approximately equal to the proton density in
the core bulk region. Here we consider a more general case that
the electron density is smaller than the proton density in core bulk
region. Actually, this is the case of
compressed, superheavy atoms in which some of the electrons have
penetrated into superheavy nuclear cores (superheavy nuclei)
\cite{Rotondo2011a, Rotondo2011b} (we call them compressed, superheavy
atoms according to Ref.~\cite{Rotondo2011b} in which a similar object was studied). 
A compressed, superheavy atom is an atom composed by 
a superheavy nuclear core (superheavy nucleus), and degenerate electrons that neutralize the system. 
Some electrons penetrate into the superheavy nuclear core and the rest surround it 
up to a distance that depends upon the compression level.
Such kind of compressed, superheavy atoms are hypothetical objects 
and could be possible to appear in the high density region 
of the neutron star crust or in other systems for example in the 
r-processes in gamma-ray bursts; studies of such kind of objects could 
provide a better understanding in the field of nuclear physics and nuclear astrophysics. In this
article, we study the surface properties of these compressed, superheavy atoms.

The article is organized as follows. In Sec.~\ref{sec:equs}, we
formulate the relativistic equations of motion for the system of
neutrons, protons and electrons fulfilling the strong and
electromagnetic interactions and $\beta$ equilibrium, and the
equations for governing the nuclear surface tension and Coulomb
energy of compressed, superheavy atoms. In Sec.~\ref{sec:NA}, we
present our discussions on the basis of the numerical analysis of
the structure, the nuclear surface tension, and the Coulomb energy
of compressed, superheavy atoms. We also apply the Bohr-Wheeler
fission mechanism \cite{Bohr1} to analyze the stability of
compressed, superheavy atoms, in Sec.~\ref{sec:NA}. We finally
give a summary in Sec.~\ref{sec:sum}. We use units with $\hbar = c =
1$ throughout the article.

%%%%%%%%%%%%%%%%%%%%%%%%%%%%%%%%%%%%%%%%%%%%%%%%%%%%%%%%%%%%%%%%
%%%%%%%%%%%%%%%%%%%%%%%%%%%%%%%%%%%%%%%%%%%%%%%%%%%%%%%%%%%%%%%%
\section{Equations of motion and surface tension}
\label{sec:equs}
%%%%%%%%%%%%%%%%%%%%%%%%%%%%%%%%%%%%%%%%%%%%%%%%%%%%%%%%%%%%%%%%
%%%%%%%%%%%%%%%%%%%%%%%%%%%%%%%%%%%%%%%%%%%%%%%%%%%%%%%%%%%%%%%%

The system of compressed, superheavy atoms under consideration is
composed of degenerate neutrons, protons, and electrons including
the strong, electromagnetic, and weak interactions and fulfilling
global charge neutrality. In this system, the electron density in
the inside bulk region ($n_{eb}$) smaller than the proton one
($n_{bp}$), i.e., $n_{eb}<n_{bp}$. We adopt the
$\sigma$-$\omega$-$\rho$ phenomenological nuclear model of Boguta
and Bodmer \cite{Boguta2} to describe the strong interactions
between the nucleons. The Lagrangian density of the model we
considered here is given by
\begin{equation}
  \mathcal{L} = \mathcal{L}_f +
  \mathcal{L}_{\sigma} + \mathcal{L}_{\omega} + \mathcal{L}_{\rho} +
  \mathcal{L}_{\gamma} + \mathcal{L}_{int},
\end{equation}
including the free-field Lagrangian densities
$\mathcal{L}_{\gamma}$, $\mathcal{L}_{\sigma}$,
$\mathcal{L}_{\omega}$, and $\mathcal{L}_{\rho}$, respectively for
the electromagnetic and the three mesonic fields, the three fermion
species (electrons, protons and neutrons) Lagrangian density
$\mathcal{L}_f$ and the interacting part $\mathcal{L}_{int}$. A
detailed description of this model can be found in
Ref.~\cite{Belvedere1}.

We adopt the compressed, superheavy atom as a spherical droplet,
so we have spherical symmetry in this system. Within the mean-field
approximation and Thomas-Fermi approximation, the equations of
motion for this system are given by
\begin{eqnarray}
  & & \frac{d^2 V}{dr^2} + \frac{2}{r} \frac{dV}{dr}
  = -4\pi e (n_p - n_e), \label{eqcom}\\
  & & \frac{d^2 \sigma}{dr^2} + \frac{2}{r} \frac{d \sigma}{dr} =
  [\partial_{\sigma} U(\sigma) + g_s n_s],
  \label{eqsig} \\
  & & \frac{d^2 \omega}{dr^2} + \frac{2}{r} \frac{d \omega}{dr} =
  -(g_{\omega} J_0^{\omega} - m_{\omega}^2 \omega),
  \label{eqom} \\
  & & \frac{d^2 \rho}{dr^2} + \frac{2}{r} \frac{d \rho}{dr} =
  -(g_{\rho} J_0^{\rho} - m_{\rho}^2 \rho),
  \label{eqrho} \\
  & & E_e^{F} = \mu_e - e V = {\rm{constant}}, \label{eqe}\\
  & & E_p^{F} = \mu_p + g_{\omega} \omega + g_{\rho} \rho
  + e V = {\rm{constant}}, \label{eqp}\\
  & & E_n^{F} = \mu_n + g_{\omega} \omega - g_{\rho} \rho  =
  {\rm{constant}}. \label{eqnn}
\end{eqnarray}
This is a special case of the EMTF system of equations \cite{Rueda1,
Belvedere1} without the presence of the gravitational interaction.
Here we have introduced the notation $\omega_0 \equiv \omega$,
$\rho_0 \equiv \rho$, and $A_0 \equiv V$ for the time components of
the meson fields, where $A$ is the electromagnetic field. $\mu_i =
\sqrt{(P_i^F)^2 + \tilde{m}_i^2}$ and $n_i = (P_i^F)^3/(3\pi^2)$ are
the free chemical potential and the number density of the
$i$-fermion species ($i=n,p,e$) with Fermi momentum $P_i^F$. The
particle effective masses are $\tilde{m}_N = m_N + g_s \sigma$ and
$\tilde{m}_e = m_e$, where $m_i$ is the rest mass of each
$i$-fermion species. $g_s$, $g_{\omega}$, and $g_{\rho}$ are the
coupling constants of the $\sigma$, $\omega$ and $\rho$ fields, and
$e$ is the fundamental electric charge. $m_{\sigma}$, $m_{\omega}$,
and $m_{\rho}$ are the masses of $\sigma$, $\omega$, and $\rho$.
$U(\sigma)$ is the scalar self-interaction potential which can be
found in e.g. Refs.~\cite{Belvedere1, RRWX2014}.

The generalized Fermi energies of electrons, protons, and neutrons,
$E_e^F$, $E_p^F$, and $E_n^F$, derived from the thermodynamic
equilibrium conditions given by the statistical physics of
multicomponent systems, are linked by the $\beta$-equilibrium
\cite{Boguta6} of protons, neutrons, and electrons,
\begin{equation} \label{betaeq}
  E_n^F = E_p^F + E_e^F.
\end{equation}

The scalar density $n_s$ is given by the expectation value
\begin{equation} \label{nsden}
  n_s = \frac{2}{(2\pi)^3}
  \sum_{i=n,p} \int_0^{P_i^F} d^3 k \frac{\tilde{m}_N}{\epsilon_i^k(k)},
\end{equation}
where $\epsilon_i^k(k) = \sqrt{k^2 + \tilde{m}^2_i}$ is the single
particle energy. In the static case, the nonvanishing components of
the currents are
\begin{eqnarray}
  J_0^{ch} &=& (n_p - n_e),\\
  J_0^{\omega} &=& n_b = (n_n + n_p),\\
  J_0^{\rho} &=& (n_p -n_n),
\end{eqnarray}
here $n_b = n_p + n_n$ is the baryon number density.

We would like to mention here that the Thomas-Fermi
approximation and the Thomas-Fermi approximation combined with the
RMFT applied to nuclei are well-known and have gained great
successes in understanding nuclear structures (see, e.g.,
Refs.~\cite{Centelles1990, Shen1998, Avancini2009}). In the present
study, we apply this approach of the Thomas-Fermi approximation
combined with RMFT to compressed, superheavy atoms, 
inspired by our new neutron star model studied in Refs.~\cite{Belvedere1, RRWX2014}. 
One of our major purposes here is to analyze the possibility of the existence 
of such ``exotic" neutron rich nuclei whose mass numbers are much larger 
than that of ordinary nuclei. Another major purpose here is to study the 
effects of the electrons and electromagnetic interaction on the surface 
properties of such a system. The study presented in this article would give us a further
understanding of the influence of the electromagnetic interaction
and electrons on the surface properties of the core-crust interface
of our new structure of neutron stars \cite{Belvedere1, RRWX2014},
hence give us a further understanding of global charge neutrality
and the structure of neutron stars.

The parameters of the nuclear model, namely the coupling constants
$g_{s}$ , $g_{\omega}$, and $g_{\rho}$, the meson masses
$m_{\sigma}$, $m_{\omega}$, and $m_{\rho}$, and the third- and
fourth- order constants of the self-scalar interactions $g_2$ and
$g_3$ are fixed by fitting nuclear experimental data, such as
saturation density, binding energy per nucleon, symmetry energy,
surface energy, and nuclear incompressibility. We here use the
parameters of the NL3 parametrization \cite{Lalazissis1} as the one
used in Refs~\cite{Belvedere1, RRWX2014}, shown in Table
\ref{tablenl3}.

\begin{table}[h]  \addtolength{\tabcolsep}{8pt}
\renewcommand{\arraystretch}{1.5}
\begin{center}
\begin{tabular}{lc||lc}
\hline\hline  $m_{\sigma}$ (MeV) & $508.194$& $g_{\omega}$       & $12.868$\\
        $m_{\omega}$ (MeV) & $782.501$& $g_{\rho}$         & $4.474$\\
        $m_{\rho}$    (MeV) & $763.000$& $g_2$ (${\rm fm}^{-1}$)  & $-10.431$\\
        $g_{s}$            & $10.217$& $g_3$              & $-28.885$\\
\hline\hline
\end{tabular}
\end{center}
\caption{The parameters of the nuclear model from NL3
\cite{Lalazissis1}.} \label{tablenl3}
\end{table}

Now we turn to the analyze of the surface tension of this system. We
construct the surface tension following a similar method in
Ref.~\cite{RRWX2014}. Since we treat the compressed, superheavy
atom as a spherical droplet, we assume a spherical surface (the size
of the system under consideration is larger than the one of ordinary
nuclei, so the nuclear curvature energy here is small compared to
the nuclear surface energy) with a small thickness separating one
finite region (inside the nuclear core region) and one semi-infinite
region (outside background region, similar to the outside crust
region in the discussion of Ref.~\cite{RRWX2014}). The number
density of the $i$-fermion species $n_i(r)$ approaches the density
of the $i$-fermion species $n_{ib}$ in the origin (the inside
region) as the position $r\rightarrow 0$, and approaches the density
in the outside region of the $i$-fermion species $n_{io}$ as the
$r\rightarrow +\infty$. To construct the surface tension, as in the
case of the semi-infinite matter model, we imagine a reference
system with sharp surfaces at radii $r_i$ ($i =
n,p,e,\sigma,\omega,\rho$) at which fermion densities and meson
fields fall discontinuously from the bulk region to the outside
region. Following a similar method of Baym-Bethe-Pethick (BBP)
\cite{Baym1}, the location of the reference surface for the
$i$-fermion species is defined by the condition that the reference
system has the same number of $i$-fermion species as the original
system,
\begin{equation} \label{sffshn}
  4\pi \int_{0}^{r_i} r^2 d r [n_i(r) - n_{ib}] + 4\pi \int_{r_i}^{\infty} r^2 d r [n_i(r) - n_{io}]
  = 0, \quad i = n,p,e. 
\end{equation}
Similar to the definition of reference surfaces for fermions, the
location of the reference surfaces for meson fields are defined by
\begin{equation} \label{sfmfshn}
  4\pi \int_{0}^{r_i} r^2 d r [F_i(r) - F_{ib}] + 4\pi \int_{r_i}^{\infty} r^2 d r [F_i(r) - F_{io}]
  = 0, \quad i = \sigma,\omega,\rho,
\end{equation}
where $F_i(r)$ is the time component of the $i$-meson field,
$F_{ib}$ is the time component of the $i$-meson field in the inside
region, and $F_{io}$ is the time component of the $i$-meson field in
the outside region.

Similar to the way of BBP \cite{Baym1}, the nuclear surface energy
can be computed as the total energy subtracting off the bulk energy,
\begin{equation}
  E_{\rm{sur}} = \sum_{i=n,p,\sigma,\omega,\rho} 4 \pi \left\{ \int_{0}^{r_i}
  r^2 [\epsilon_{i}(r) - \epsilon_{ib}] dr
  + \int_{r_i}^{\infty} r^2 [\epsilon_{i}(r) - \epsilon_{io}] dr
  \right\},
\end{equation}
and the Coulomb energy is
\begin{equation}
  E_{\rm{coul}} = 4\pi \int_{0}^{\infty} r^2 \epsilon_E(r) dr,
\end{equation}
where $\epsilon_i (r)$ is the energy density of the $i$ species of
fermion or meson fields, $\epsilon_{ib}$ is the energy density of
the $i$ species of fermion or meson fields in the center of the
system (the inside region), $\epsilon_{io}$ is the energy density of
the $i$ species of fermion or meson field in the outside region, and
$\epsilon_E (r)$ is the energy density of the electric field.
Similar to the energy densities given in Ref.~\cite{RRWX2014}, the
energy density of the $i$-fermion species $\epsilon_i (r)$ is
\begin{eqnarray}
  \epsilon_i (r)
  &=& \frac{1}{8\pi^2} \bigg\{ P_i^F \sqrt{(P_i^F)^2 + \tilde{m}_i^2} \left[2(P_i^F)^2 +
  \tilde{m}_i^2\right] \nonumber\\
  & & \qquad - \tilde{m}^4 \ln\frac{P_i^F + \sqrt{(P_i^F)^2 + \tilde{m}_i^2}}{\tilde{m}_i}
  \bigg\},  \label{edf}
\end{eqnarray}
and the energy densities of the meson fields in this spherical
system are
\begin{eqnarray} \label{edmfr}
  \epsilon_{\sigma}(r)&=& \frac{1}{2} \bigg( \frac{d\sigma}{dr}
  \bigg)^2 + U(\sigma),\\
  \epsilon_{\omega}(r)&=& \frac{1}{2} \bigg( \frac{d\omega}{dr}
  \bigg)^2 + \frac{1}{2} m_{\omega}^2 \omega^2,\\
  \epsilon_{\rho}(r)&=& \frac{1}{2} \bigg( \frac{d\rho}{dr}
  \bigg)^2 + \frac{1}{2} m_{\rho}^2 \rho^2,\\
  \epsilon_{E}(r)&=& \frac{1}{8\pi} \bigg( \frac{dV}{dr}
  \bigg)^2.
\end{eqnarray}

The nuclear surface tension is given as the nuclear surface energy
per unit area,
\begin{equation} \label{stns}
  \sigma_{\rm{Ns}} = \frac{E_{\rm{sur}}}{4\pi r_n^2},
\end{equation}
and similarly we obtain the Coulomb energy per unit area (the
surface tension for the electric field)
\begin{equation} \label{stcs}
  \sigma_{\rm{Cs}} = \frac{E_{\rm{coul}}}{4\pi r_n^2},
\end{equation}
where $r_n$ is the reference radius of neutrons defined by
Eq.~(\ref{sffshn}). Since the neutron number is much larger than the
proton number in the system, so it is reasonable to set the radius
of neutrons to be the radius of the nucleus to estimate the surface
tensions; this is consistent with the existence of the neutrons halo
or neutron skin effect \cite{Tamii1}.

The relation between the nuclear surface energy and the Coulomb
energy is very important for a nucleus. As shown by Bohr and Wheeler
\cite{Bohr1} when the condition
\begin{equation} \label{bwc}
  E_{\rm{coul}} > 2E_{\rm{sur}}
\end{equation}
is satisfied, the nucleus becomes unstable against nuclear fission.
A careful analysis on the derivation of this condition shows that
the Bohr-Wheeler condition given by Eq.~(\ref{bwc}) applies also to
our system \cite{RRWX2014}.

%%%%%%%%%%%%%%%%%%%%%%%%%%%%%%%%%%%%%%%%%%%%%%%%%%%%%%%%%%%%%%%%%%%%%%
%%%%%%%%%%%%%%%%%%%%%%%%%%%%%%%%%%%%%%%%%%%%%%%%%%%%%%%%%%%%%%%%%%%%%%
\section{Numerical analysis}
\label{sec:NA}
%%%%%%%%%%%%%%%%%%%%%%%%%%%%%%%%%%%%%%%%%%%%%%%%%%%%%%%%%%%%%%%%%%%%%%
%%%%%%%%%%%%%%%%%%%%%%%%%%%%%%%%%%%%%%%%%%%%%%%%%%%%%%%%%%%%%%%%%%%%%%

Following a similar procedure in Refs.~\cite{Belvedere1, RRWX2014},
we can solve the equations (\ref{eqcom})-(\ref{eqnn}) together with
the $\beta$-equilibrium (\ref{betaeq}) to obtain the fermion-density
and meson-field profiles. This system of equations can be
numerically solved with appropriate boundary conditions and
approximations, as shown in Refs.~\cite{Belvedere1, RRWX2014}.

In order to obtain a solution of these equations, we set a value for
the baryon number density $n_{\rm{bb}} = n_{nb}+n_{pb}$ in the
region near the center, and we set a small electron density $n_{eb}
= y_e n_{pb}$ in the region near the center with electron fraction
$y_e<1$. As described in Refs.~\cite{Belvedere1, RRWX2014}, the
fermion densities $n_{io}$ in the outside region depend on the
density at the base of the background under consideration (similar
to the crust in the discussion of Ref.~\cite{RRWX2014}). The
background matter is composed of a nuclei lattice in a background of
degenerate electrons, whose density is denoted here as
$n_{e}^{\rm{bg}}$. In addition, there are free neutrons in the
background when the density $\rho_{\rm{bg}}$ of the background is
higher than the neutron-drip density $\rho_{\rm{drip}} \approx 4.3
\times 10^{11}$ g cm$^{-3}$ \cite{Baym1}. So when the density
$\rho_{\rm{bg}}$ of the background is smaller than the neutron-drip
density $\rho_{\rm{drip}}$, i.e., $\rho_{\rm{bg}} <
\rho_{\rm{drip}}$, we set the proton density and the neutron density
to zero in the outside region while the electron density matches the
value $n_{e}^{\rm{bg}}$ of the density of background electrons,
i.e., $n_{eo} = n_{e}^{\rm{bg}}$. When $\rho_{\rm{bg}}
> \rho_{\rm{drip}}$ both the neutron density and the electron density have to match
their corresponding background densities, i.e., $n_{eo} =
n_{e}^{\rm{bg}}$ and $n_{no} = n_{n}^{\rm{bg}}$, where
$n_n^{\rm{bg}}$ is the neutron density in the background. As shown
in Ref.~\cite{Baym1} there is no proton drip in the systems under
consideration, so we keep the outside proton density value as zero.
In order to set the matching density values for electrons and
neutrons we use the relation between the free neutron density and
the electron density in Section 6 of Ref.~\cite{Baym1}.

\begin{figure}[h]
\begin{center}
\includegraphics[width=\columnwidth]{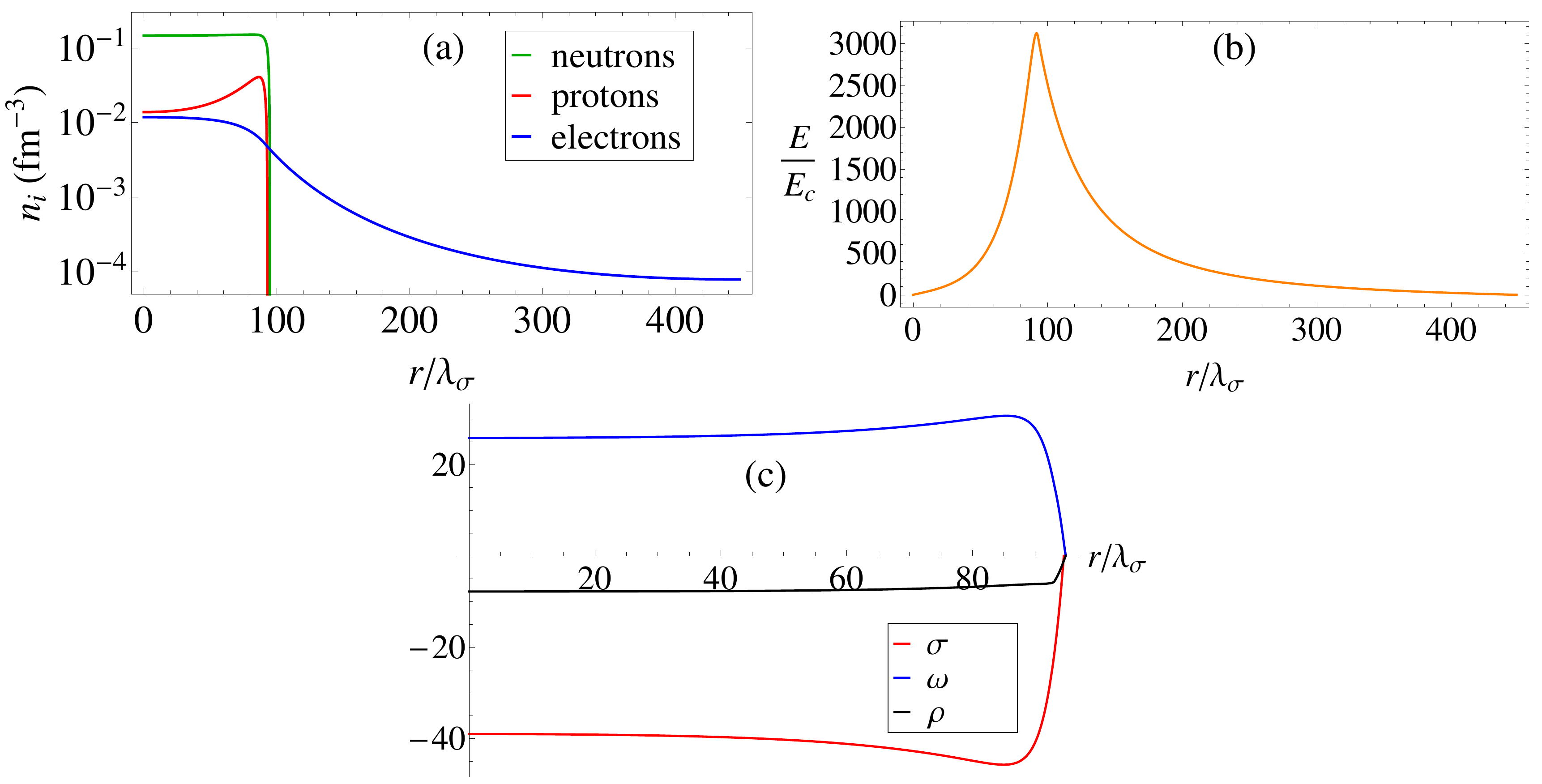}
\caption{(Color online) (a): fermion density profiles in units of
$\rm{fm^{-3}}$. (b): electric field in units of the critical field
$E_c$. (c): meson fields $\sigma$, $\omega$, and $\rho$ in units of
$\rm{MeV}$. Here we set $P^{F}_{\rm{eb}} = 0.95 P^{F}_{\rm{pb}}$,
the baryon number density in the region near the center is the
nuclear saturation density $n_{\rm{nucl}}$, and the density in the
outside (background) region is the neutron-drip density
$\rho_{\rm{bg}} = \rho_{\rm{drip}} \approx 4.3\times 10^{11}$ g
cm$^{-3}$. $\lambda_{\sigma} = \hbar /(m_{\sigma}c) \sim 0.4$ ${\rm
fm}$ is the Compton wavelength of the $\sigma$ meson.} \label{ne95}
\end{center}
\end{figure}

\begin{figure}[h]
\begin{center}
\includegraphics[width=\columnwidth]{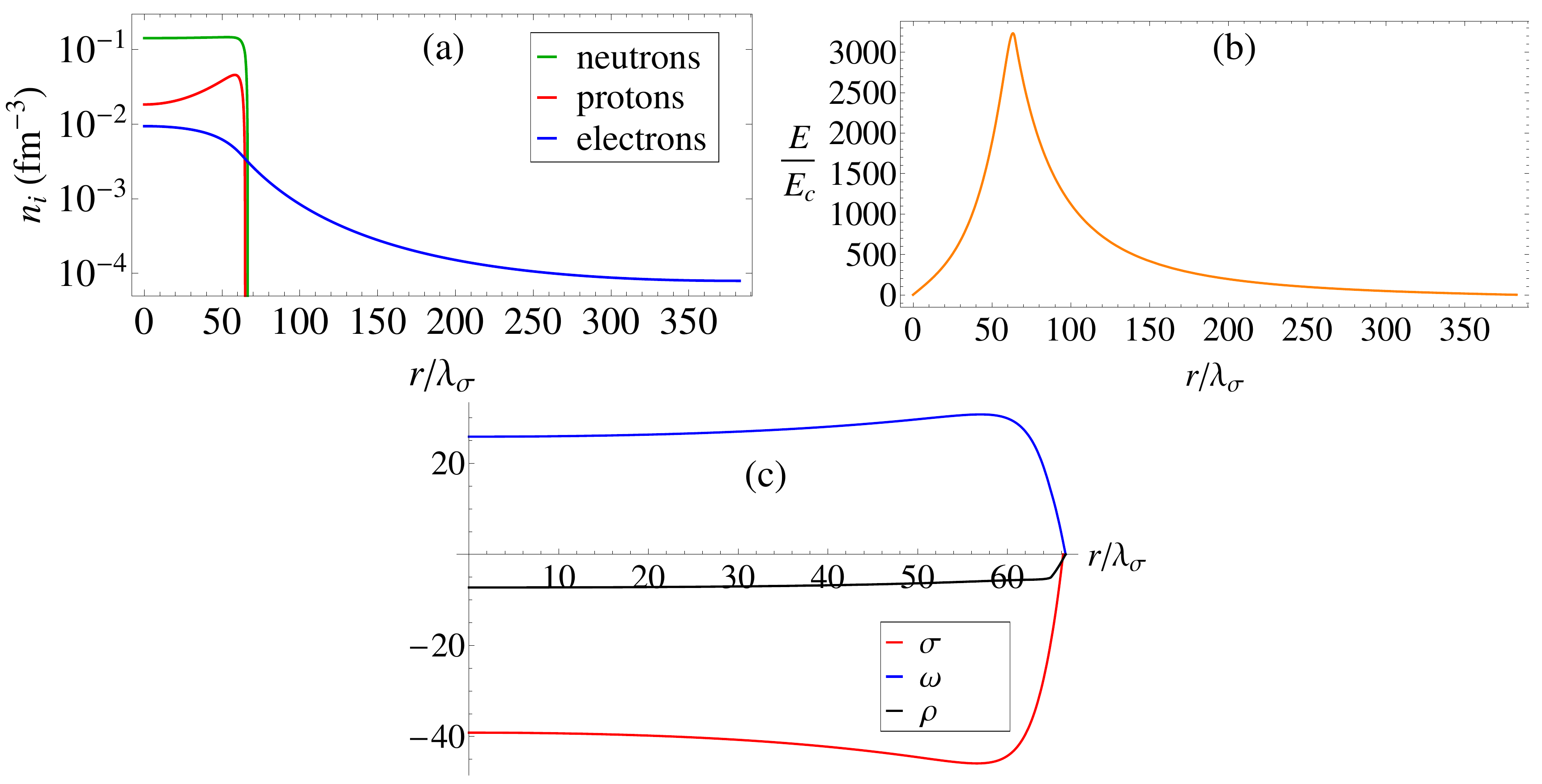}
\caption{(Color online) (a): fermion density profiles in units of
$\rm{fm^{-3}}$. (b): electric field in units of the critical field
$E_c$. (c): meson fields $\sigma$, $\omega$, and $\rho$ in units of
$\rm{MeV}$. Here we set $P^{F}_{\rm{eb}} = 0.8 P^{F}_{\rm{pb}}$, the
baryon number density in the region near the center is the nuclear
saturation density $n_{\rm{nucl}}$, and the density in the outside
(background) region is the neutron-drip density $\rho_{\rm{bg}} =
\rho_{\rm{drip}} \approx 4.3\times 10^{11}$ g cm$^{-3}$.}
\label{ne80}
\end{center}
\end{figure}

\begin{figure}[h]
\begin{center}
\includegraphics[width=\columnwidth]{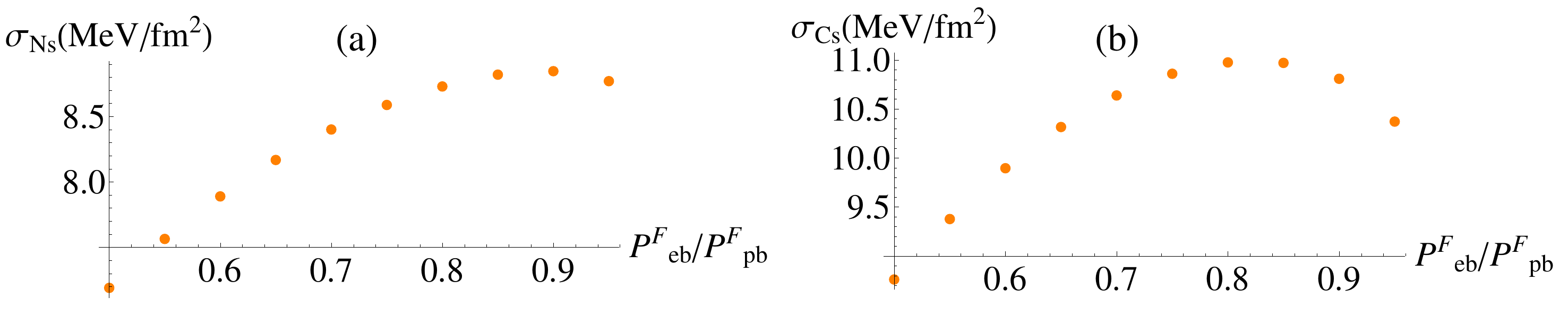}
\caption{(Color online) The dependence of the surface tension on the
ratio $P^{F}_{eb}/P^{F}_{pb}$.  The baryon number density in the
region near the center is the nuclear saturation density
$n_{\rm{nucl}}$, and the fermion densities and meson fields tend to
be zero in the outside region. (a): nuclear surface tension
$\sigma_{Ns}$. (b): Coulomb energy per unit area $\sigma_{Cs}$.}
\label{stdne}
\end{center}
\end{figure}

\begin{figure}[h]
\begin{center}
\includegraphics[width=\columnwidth]{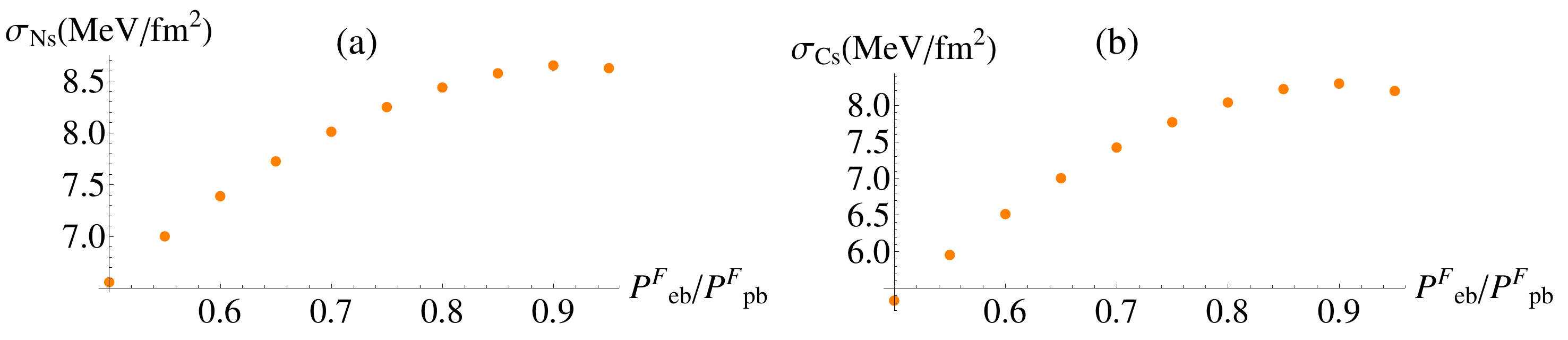}
\caption{(Color online) The dependence of the surface tension on the
ratio $P^{F}_{\rm{eb}}/P^{F}_{\rm{pb}}$. The baryon number density
in the region near the center is the nuclear saturation density
$n_{\rm{nucl}}$, and the density in the outside region is the
neutron-drip density $\rho_{\rm{bg}} = \rho_{\rm{drip}} \approx
4.3\times 10^{11}$ g cm$^{-3}$. (a): nuclear surface tension
$\sigma_{\rm{Ns}}$. (b): Coulomb energy per unit area
$\sigma_{\rm{Cs}}$.} \label{stdnend}
\end{center}
\end{figure}

\begin{figure}[h]
\begin{center}
\includegraphics[width=\columnwidth]{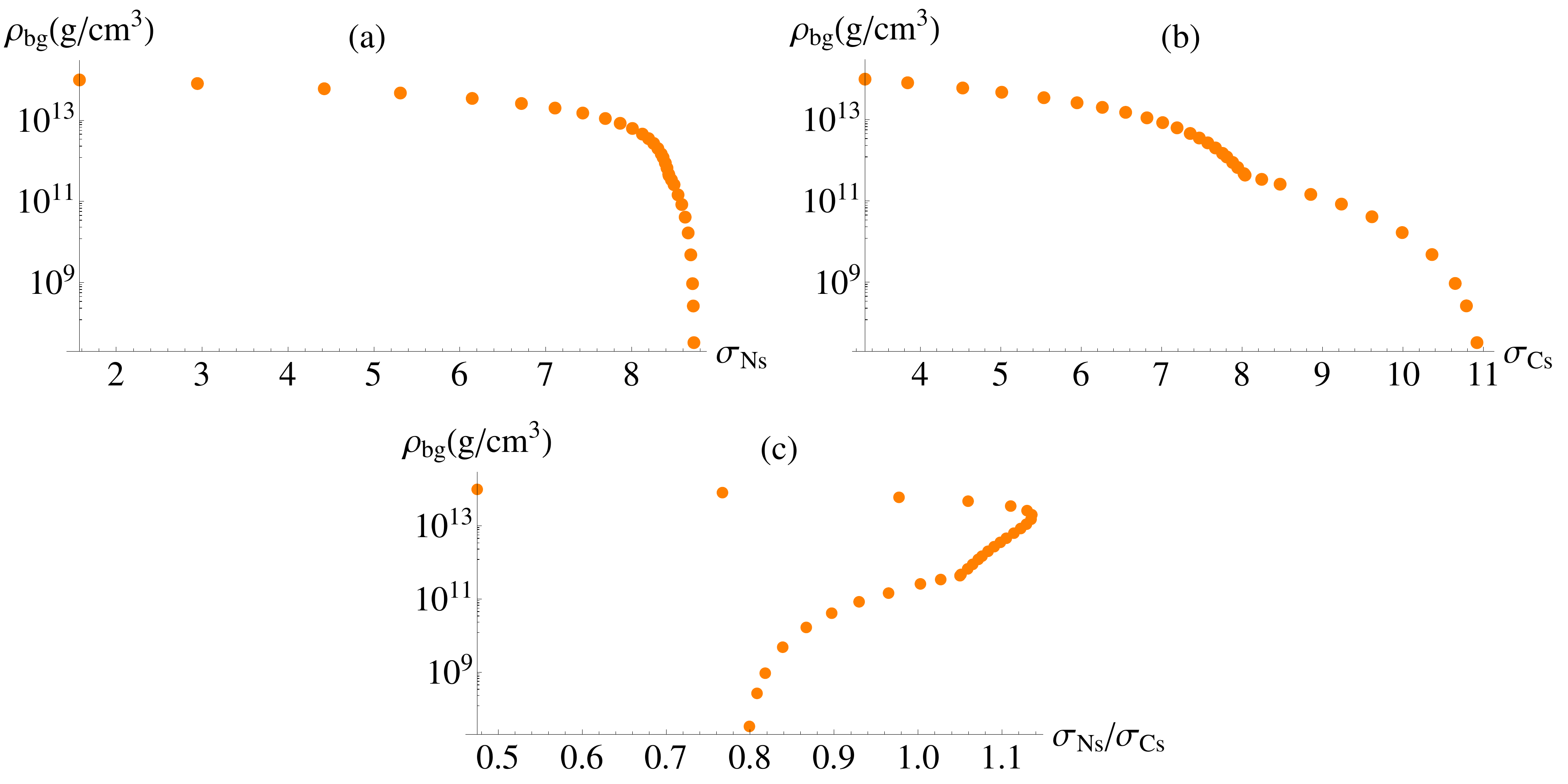}
\caption{(Color online) The dependence of the surface tension on the
density $\rho_{\rm{bg}}$ of the background. Here we set
$P^{F}_{\rm{eb}} = 0.8 P^{F}_{\rm{pb}}$ and the baryon number
density in the region near the center is the nuclear saturation
density $n_{\rm{nucl}}$. (a): nuclear surface tension
$\sigma_{\rm{Ns}}$, in units of MeV fm$^{-2}$. (b): Coulomb energy
per unit area $\sigma_{\rm{Cs}}$, in units of MeV fm$^{-2}$. (c)
Ratio of the nuclear surface tension and the Coulomb energy per unit
area, $\sigma_{\rm{Ns}}/\sigma_{\rm{Cs}}$.} \label{stne80nd}
\end{center}
\end{figure}

\begin{figure}[h]
\begin{center}
\includegraphics[width=\columnwidth]{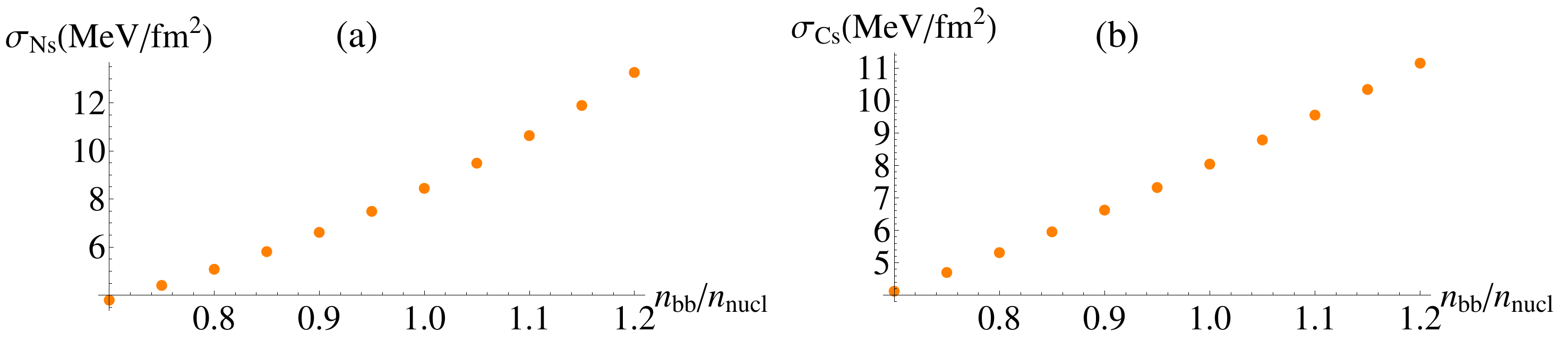}
\caption{(Color online) The dependence of the surface tension on the
baryon number density in the region near the center ($n_{\rm{bb}}$).
Here $P^{F}_{\rm{eb}} = 0.8 P^{F}_{\rm{pb}}$, and the density in the
outside region is the neutron-drip density $\rho_{\rm{bg}} =
\rho_{\rm{drip}} \approx 4.3\times 10^{11}$ g cm$^{-3}$. (a):
nuclear surface tension $\sigma_{\rm{Ns}}$. (b): Coulomb energy per
unit area $\sigma_{\rm{Cs}}$.} \label{st08nednb}
\end{center}
\end{figure}

As shown in Refs.~\cite{Belvedere1, RRWX2014}, the
transition interface that we are interesting in appears near the
nuclear saturation density $n_{\rm{nucl}} = 0.16$ $\rm{fm^{-3}}$. In
order to study the compressed, superheavy atoms and 
the influence of the electrons and electromagnetic
interaction on the surface properties of the system, we assume at
first the baryon number density in the region near the center to be
the nuclear saturation density (results presented in
Figs.~\ref{ne95}-\ref{stne80nd}), i.e., $n_{\rm{bb}} = n_{\rm{nucl}}
= 0.16$ $\rm{fm^{-3}}$. At the end of this section, we will also
study the influence of baryon number density (results presented in
Fig.~\ref{st08nednb} and Table \ref{tablesnbdne}).

The results of the solutions of two examples are shown in
Fig.~\ref{ne95} for the case $P^{F}_{\rm{eb}} = 0.95
P^{F}_{\rm{pb}}$ and in Fig.~\ref{ne80} for the case
$P^{F}_{\rm{eb}} = 0.8 P^{F}_{\rm{pb}}$, when the density in the
outside (background) region is the neutron-drip density
$\rho_{\rm{bg}} = \rho_{\rm{drip}} \approx 4.3\times 10^{11}$ g
cm$^{-3}$. We have introduced the notations $P^{F}_{\rm{eb}}$ for
the Fermi momentum of electrons in the region near the center of the
system, and $P^{F}_{\rm{pb}}$ for the Fermi momentum of protons in
the region near the center of the system. It is also worth
mentioning here that the typical mass number of these compressed, superheavy
atoms is $\sim 10^4$; e.g., $A\sim 35000$ and $Z/A \sim
0.154$ for the case shown in Fig.~\ref{ne95}, and $A\sim 12000$ and
$Z/A \sim 0.189$ for the case shown in Fig.~\ref{ne80}, where $A$ is 
the total number of nucleons (mass number) and $Z$ is the total number of protons. 
The mass numbers of these compressed, superheavy atoms are 
much larger than that of ordinary nuclei.

As shown in Fig.~\ref{ne95}, when the difference between the
electron density and the proton density in the region near the
center of the system ($n_{pb}-n_{eb}$) is small, the fermion-density
and meson-field profiles are similar to their counterparts in the
case of semi-infinite matter (electron density nearly equal to the
proton density in the inside bulk region $n_{eb} \simeq n_{pb}$).
Comparing to the results in the case of the electron density being
approximately equal to the proton density in the core bulk region
shown in Ref.~\cite{RRWX2014}, the bump of the proton profile is
larger in this case, as expected from the fact that the internal
electric field is less screened than the case of $n_{eb} \simeq
n_{pb}$. We can also see from Figs.~\ref{ne95}-\ref{ne80}, how the
fermion and meson-field profiles change for increasing charge
separations ($n_{pb}-n_{eb}$).

Using the definitions in Eqs.~(\ref{stns}) and (\ref{stcs}), we
obtain the surface tensions for compressed, superheavy atoms. The
dependence of the surface tension on the ratio of the electron Fermi
momentum and the proton Fermi momentum in the region near the center
of the system ($P^{F}_{\rm{eb}}/P^{F}_{\rm{pb}}$) is shown in
Fig.~\ref{stdne} for the case of the fermion densities and meson
fields tending to be zero in the outside region, and
Fig.~\ref{stdnend} for the case of the density in the outside
(background) region is the neutron-drip density $\rho_{\rm{bg}} =
\rho_{\rm{drip}} \approx 4.3\times 10^{11}$ g cm$^{-3}$. From the
results, the system is stable with respect to the Bohr-Wheeler
condition (\ref{bwc}) of the stability, in all ratios
$P^{F}_{\rm{eb}}/P^{F}_{\rm{pb}}$ we consider. This is the result of
the penetration of the relativistic electrons into the nucleus (see
also Refs.~\cite{Rotondo2011a, Rotondo2011b}). This in principle implies the 
possibility of the existence of such kind of compressed, superheavy atoms. As shown in
Fig.~\ref{stdne}, the nuclear surface tension $\sigma_{Ns}$ first
increases and then decreases when the difference between the
electron density and the proton density increases, and the nuclear
surface tension tends to the phenomenological result ($\sim 1$ MeV
fm$^{-2}$) without the presence of electrons in the inside bulk
region studied in the nuclear physics\cite{Baym1}. There are two
effects which influence on the nuclear surface tension
$\sigma_{Ns}$: (I) for $n_{eb}<n_{pb}$ the bump of the proton
profile around the nuclear surface changes as shown in
Figs.~\ref{ne95}--\ref{ne80}, and (II) the higher the difference
($n_{pb}-n_{eb}$) is, the lower the nuclear asymmetry. As a
consequence, the total energy of the system decreases. The
combination of these two effects leads to the results of the nuclear
surface tension $\sigma_{Ns}$ shown in Fig.~\ref{stdne}.

Comparing the results of Fig.~\ref{stdne} and Fig.~\ref{stdnend}, we
can find that the electrons in the outside region have strong
effects on the surface structure of compressed, superheavy atoms
considered here. The increase of the electron density in the outside
region effectively reduces the Coulomb energy per unit area
$\sigma_{\rm{Cs}}$, as well as the nuclear surface tension
$\sigma_{\rm{Ns}}$. This effect is enhanced when increasing
difference between the electron density and the proton density in
the region near the center of the system ($n_{pb}-n_{eb}$), as shown
in Figs.~\ref{stdne}-\ref{stdnend}. This effect is mainly due to the
reason that the electrons have a strong influence on the bump on the
profiles, leading to a strong effect on the surface structure and
the surface tensions $\sigma_{\rm{Ns}}$ and $\sigma_{\rm{Cs}}$.
These results provide the evidence of strong effects of the
electromagnetic interaction and electrons on structure of the
system. This result of the effect due to the electrons in the
outside region as shown by the comparison of Fig.~\ref{stdne} and
Fig.~\ref{stdnend} is different from the case studied in
Ref.~\cite{RRWX2014} where the electron density in the inside bulk
region ($n_{eb}$) is nearly equal to the proton one ($n_{bp}$). In
the case shown in Ref.~\cite{RRWX2014}, the effect of the electrons
in the outside region is small when the density $\rho_{\rm{bg}}$ in
the outside region is smaller than the neutron-drip density,
$\rho_{\rm{bg}} < \rho_{\rm{drip}}$.

We now turn to study the effect of the free neutrons in the
background (the outside region) on the surface properties of
compressed, superheavy atoms. The dependence of the surface
tension on the density $\rho_{\rm{bg}}$ of the background for the
case of $P^{F}_{\rm{eb}} = 0.8 P^{F}_{\rm{pb}}$ is shown in
Fig.~\ref{stne80nd}. As shown in Fig.~\ref{stne80nd}(c), the
Bohr-Wheeler condition (\ref{bwc}) for the instability is reached at
a background density $\rho_{\rm{bg}}^{\rm{crit}} \sim $ $9.7 \times
10^{13}$ g cm$^{-3}$, so the system becomes unstable against fission
when $\rho_{\rm{bg}}>\rho_{\rm{bg}}^{\rm{crit}}$. This imposes a
physical upper limit to the density of the background for
compressed, superheavy atoms with $P^{F}_{\rm{eb}} = 0.8
P^{F}_{\rm{pb}}$. This critical background density
$\rho_{\rm{bg}}^{\rm{crit}}$ is smaller than the one for the case of
the electron density in the inside bulk region being nearly equal to
the proton one ($n_{eb}\simeq n_{bp}$) discussed in
Ref.~\cite{RRWX2014}. This implies that the difference between the
electron density and the proton density in the region near the
center of the system ($n_{pb}-n_{eb}$) can decrease the stability of
compressed, superheavy atoms.

The results in Fig.~\ref{stne80nd} clearly show the strong effect of
the fermions in the outside (background) region on the surface
structure of compressed, superheavy atoms, as we have discussed
above in the comparison of Fig.~\ref{stdne} and Fig.~\ref{stdnend}.
The Coulomb energy per unit area $\sigma_{\rm{Cs}}$ and the nuclear
surface tension $\sigma_{\rm{Ns}}$ change significantly as changing
the density $\rho_{\rm{bg}}$ of the background (the outside region),
in both cases: (I) the density $\rho_{\rm{bg}}$ of the background is
higher than the neutron-drip density $\rho_{\rm{drip}}$; (II) the
density $\rho_{\rm{bg}}$ of the background is smaller than the
neutron-drip density $\rho_{\rm{drip}}$.

In the previous discussions, we have assumed the baryon
number density in the region near the center to be the nuclear
saturation density $n_{\rm{nucl}}$ in symmetric matter, to study the
influence of the electrons on the surface properties of the
transition interface \cite{Belvedere1, RRWX2014}. However, the
saturation density in nuclei can be different while changing the
asymmetry parameter (see, e.g., Refs.~\cite{Papakonstantinou2013,
Aymard2014}). Therefore, it would be necessary to analyze the
influence of the baryon number density on the surface tensions. The
dependence of the surface tension on the baryon number density in
the region near the center ($n_{\rm{bb}}$) is show in
Fig.~\ref{st08nednb}, for the case of $P^{F}_{\rm{eb}} = 0.8
P^{F}_{\rm{pb}}$ and $\rho_{\rm{bg}} = \rho_{\rm{drip}} \approx
4.3\times 10^{11}$ g cm$^{-3}$. Comparing with the results in
Ref.~\cite{RRWX2014}, the dependence of the surface tension on the
baryon number density shown in Fig.~\ref{st08nednb} for the case of
compressed, superheavy atoms has a similar behavior as in the case
discussed in Ref.~\cite{RRWX2014} for the core-crust interface of
neutron stars ($n_{eb} \approx n_{pb}$). Therefore, we can conclude
that the effects of the baryon number density on the surface
tensions $\sigma_{\rm{Ns}}$ and $\sigma_{\rm{Cs}}$ for the case of
compressed, superheavy atoms are similar to the ones for the case
the core-crust interface of neutron stars ($n_{eb} \approx n_{pb}$)
\cite{RRWX2014}.

\begin{table}[h] \addtolength{\tabcolsep}{8pt}
\renewcommand{\arraystretch}{1.5}
\begin{center}
\begin{tabular}{l c c c}
\hline\hline $P^{F}_{\rm{eb}}/P^{F}_{\rm{pb}}$ & $\sigma_{\rm{Ns}}$
& $\sigma_{\rm{Cs}}$ & $E_{\rm{sur}}/A^{2/3}$\\
\hline  $0.5$     & $3.84$     & $3.57$ & $69.7$\\
        $0.8$     & $5.07$     & $5.31$ & $90.4$\\
        $0.95$    & $5.17$     & $5.36$ & $93.1$\\
\hline\hline
\end{tabular}
\end{center}
\caption{Surface tensions $\sigma_{\rm{Ns}}$ and $\sigma_{\rm{Cs}}$
in MeV fm$^{-2}$ and nuclear surface energy per surface nucleon
$E_{\rm{sur}}/A^{2/3}$ in MeV for selected values of
$P^{F}_{\rm{eb}}/P^{F}_{\rm{pb}}$. Here the density in the outside
region is the neutron-drip density $\rho_{\rm{bg}} =
\rho_{\rm{drip}} \approx 4.3\times 10^{11}$ g cm$^{-3}$, and the
baryon number density in the region near the center ($n_{\rm{bb}}$)
is equal to $0.8 n_{\rm{nucl}}$.} \label{tablesnbdne}
\end{table}

Furthermore, we show in Table \ref{tablesnbdne} the
surface tensions of compressed, superheavy atoms for selected
values of $P^{F}_{\rm{eb}}/P^{F}_{\rm{pb}}$ when a smaller baryon
number density in the region near the center is adopted
($n_{\rm{bb}} = 0.8 n_{\rm{nucl}}$). We can learn from Table
\ref{tablesnbdne} and Fig.~\ref{stdnend} that the dependence of the
surface tension on the ratio $P^{F}_{\rm{eb}}/P^{F}_{\rm{pb}}$ for
the case of compressed, superheavy atoms with a smaller baryon
number density in the region near the center has a similar behavior
as in the case when the baryon number density in the region near the
center is $n_{\rm{nucl}}$.

It is worth mentioning that the properties of
medium-mass and heavy nuclear clusters embedded in a gas of nucleons
were analyzed in Refs.~\cite{Papakonstantinou2013, Aymard2014}. The
calculations varying the cluster size and isospin asymmetry over a
large domain of $N$ (neutron number) and $Z$ covering the whole
periodic table well beyond the neutron drip line, were performed.
The nuclear surface energy per surface nucleon
$E_{\rm{sur}}/A^{2/3}$ obtained in Refs.~\cite{Papakonstantinou2013,
Aymard2014} is in the order of $~20$ MeV (the value depends on the
parameters such as the asymmetry of the nucleus and the density of
the nucleon gas \cite{Papakonstantinou2013, Aymard2014}). Comparing
the result shown in Table \ref{tablesnbdne} and the result obtained
in Refs.~\cite{Papakonstantinou2013, Aymard2014}, compressed, superheavy
atoms under consideration have larger nuclear surface
energies per surface nucleon $E_{\rm{sur}}/A^{2/3}$. This is mainly
due to the fact that the electromagnetic interaction and the
presence of electrons change the proton and neutron density
profiles, as we have discussed in Ref.~\cite{RRWX2014}. As we have
shown in Ref.~\cite{RRWX2014}, the nuclear surface tension we
obtained for the case without the presence of electrons matches the
result in literature for ordinary nuclear matter. The trend from
compressed, superheavy atoms to ordinary nuclei is also shown in
Figs.~\ref{stdne}-\ref{stdnend} and Table \ref{tablesnbdne} when
reducing the electron density.

%%%%%%%%%%%%%%%%%%%%%%%%%%%%%%%%%%%%%%%%%%%%%%%%%%%%%%%%%%%%%%%%
%%%%%%%%%%%%%%%%%%%%%%%%%%%%%%%%%%%%%%%%%%%%%%%%%%%%%%%%%%%%%%%%
\section{Summary}
\label{sec:sum}
%%%%%%%%%%%%%%%%%%%%%%%%%%%%%%%%%%%%%%%%%%%%%%%%%%%%%%%%%%%%%%%%
%%%%%%%%%%%%%%%%%%%%%%%%%%%%%%%%%%%%%%%%%%%%%%%%%%%%%%%%%%%%%%%%

Following our study \cite{RRWX2014} of the surface properties of the
core-crust interface of neutron stars with global charge neutrality,
we study the surface properties of compressed, superheavy atoms.
By compressed, superheavy atom we mean an atom composed by 
a superheavy nuclear core (superheavy nucleus) with mass number 
of the order of $10^4$, and degenerate electrons that neutralize the system. 
Some electrons penetrate into the superheavy nuclear core and the rest surround it 
up to a distance that depends upon the compression level. We have adopted
both the Thomas-Fermi approximation and RMFT approach and taken into
account the strong, weak, and electromagnetic interactions. We
numerically studied the structure of compressed, superheavy atoms,
computed the nuclear surface tension and Coulomb energy of
compressed, superheavy atoms, and analyzed the influence of the
electron component and the background matter on the properties of
these compressed, superheavy atoms.

We assume at
first the baryon number density in the region near the center to be
the nuclear saturation density $n_{\rm{nucl}}$ as in
Ref.~\cite{RRWX2014}. We show how the nuclear surface tension
$\sigma_{Ns}$ and the Coulomb energy per unit area
$\sigma_{\rm{Cs}}$ are drastically affected by the decreasing of
electron to proton density ratio in the region near the center of
compressed, superheavy atoms (see Figs.~\ref{ne95}, \ref{ne80},
\ref{stdne}, and \ref{stdnend}). This is due to the increasing of
proton repulsion and the decreasing of nuclear asymmetry when
decreasing electron to proton density ratio in the region near the
center of compressed, superheavy atoms. If the charge separation
is small (i.e., the electron density $n_{eb}$ in the inside region
is slightly smaller than the proton one $n_{pb}$; it means most of
electrons penetrate into nuclear cores), the surface properties are
closed to the ones discussed in Ref.~\cite{RRWX2014} for the
core-crust interface of neutron stars ($n_{eb} \approx n_{pb}$). If
the charge separation is large (i.e., the electron density in the
inside region is much smaller than the proton one $n_{pb}$; it means
only some of electrons penetrate into nuclear cores), the surface
properties approach to the results without the presence of electrons
inside nuclei, studied in the nuclear physics.

It is also shown (see Figs.~\ref{stdne}, \ref{stdnend}, and
\ref{stne80nd}) that electrons in the outside (background) region
have strong effects on the surface properties of compressed, 
superheavy atoms. The increase of the electron density in the
outside region effectively reduces the Coulomb energy per unit area
$\sigma_{\rm{Cs}}$ and nuclear the surface tension
$\sigma_{\rm{Ns}}$ even if the density $\rho_{\rm{bg}}$ of the
background (the outside region) is smaller than the neutron-drip
density $\rho_{\rm{drip}}$. This effect is enhanced when increasing
difference between the electron density and the proton density in
the region near the center of the system ($n_{pb}-n_{eb}$) (the
inside region). These results show the evidence of strong effects of
the electromagnetic interaction and electrons on the structure of
compressed, superheavy atoms.

Base on the above numerical results, we studied the instability of
compressed, superheavy atoms against Bohr-Wheeler surface
deformations. We find that the instability sets in at a critical
density of the background $\rho_{\rm{bg}}^{\rm{crit}} \sim $ $9.7
\times 10^{13}$ g cm$^{-3}$ for compressed, superheavy atoms with
$P^{F}_{\rm{eb}} = 0.8 P^{F}_{\rm{pb}}$. This critical background
density $\rho_{\rm{bg}}^{\rm{crit}}$ is smaller than the one
obtained in Ref.~\cite{RRWX2014}, where the electron density in the
inside bulk region is nearly equal to the proton one ($n_{eb}\simeq
n_{bp}$). This implies that the stability of the system can be
decreased by increasing difference between the electron density and
the proton density in the region near the center of compressed, superheavy atoms ($n_{pb}-n_{eb}$).

We also studied the influence of the baryon number
density on the nuclear surface tension and the Coulomb energy per
unit area of compressed, superheavy atoms. The results show that
the effects of the baryon number density on the surface tensions
$\sigma_{\rm{Ns}}$ and $\sigma_{\rm{Cs}}$ for the case of
compressed, superheavy atoms are similar to the ones for the case
the core-crust interface of neutron stars ($n_{eb} \approx n_{pb}$)
\cite{RRWX2014}.

We showed through the Bohr-Wheeler condition, 
the possibility of the existence of compressed, superheavy 
atoms with $A$ in the order of $10^4$. The mass number of 
such kind of ``exotic" neutron rich nuclei is about one order of 
magnitude larger than the usual neutron rich nuclei of the mass 
number being usually up to the order of $10^3$, studied in various
models such as pasta structures (see, e.g., 
Refs.~\cite{Grill1, Bao2014, Bao2015, Newton2009, Schuetrumpf2015}) and heavy 
nuclear clusters (see, e.g., Refs.~\cite{Papakonstantinou2013, Aymard2014}). 
Such kind of compressed, superheavy atoms could be possible to appear in 
the high density region of the neutron star crust or in the r-processes in gamma-ray bursts,
since their existence is possible according to the Bohr-Wheeler condition as discussed in the present paper. 

The results of this work show the effects of the
electrons and electromagnetic interaction on the surface properties
of the system composed of degenerate neutrons, protons, and
electrons fulfilling global charge neutrality. This would give us a
further understanding of the core-crust interface of our new
structure of neutron stars analyzed in Refs.~\cite{Belvedere1,
RRWX2014}.

To end this article, we would like to mention that another kind of
instability in nuclear matter, corresponding to the transition
density from nonuniform to uniform nuclear matter, are widely
discussed in the literature (see, e.g., Refs.~\cite{Oyamatsu2007,
Avancini1, Grill1, Pais2016, Horowitz2001, Li2008}). When the density reaches
this transition density, the pasta structures become unstable and
are dissolved into uniform matter. The transition density from
nonuniform to uniform nuclear matter is around $\sim 0.08$
$\rm{fm}^{-3}$, and strongly depends on approach to obtain it; it
can vary from $\sim 0.1$ $\rm{fm}^{-3}$ to $\sim 0.05$
$\rm{fm}^{-3}$ in different parameters of nuclear model (see, e.g.,
Refs.~\cite{Oyamatsu2007, Avancini1, Grill1, Pais2016, Horowitz2001, Li2008}).
This transition density from nonuniform to uniform nuclear matter is
in the same order of the instability (critical) density obtained in
the present article (baryon number density $\sim 0.05$
$\rm{fm}^{-3}$ for the case of $P^{F}_{\rm{eb}} = 0.8
P^{F}_{\rm{pb}}$) and in Ref.~\cite{RRWX2014} (baryon number density
$\sim 0.07$ $\rm{fm}^{-3}$ for the case of $n_{eb}\simeq n_{bp}$
presented in Ref.~\cite{RRWX2014}). It would be interesting to
compare and contrast the instability mechanism analyzed in the
present article and in Ref.~\cite{RRWX2014} with the one of the
transition density from nonuniform to uniform nuclear matter in the
literature, and analyze the difference and links between these two
instability mechanisms. However, these studies are out of the scope
of this work and we leave these studies for future work.

\ack{Yuan-Bin Wu acknowledges the support given by the Erasmus Mundus
Joint Doctorate Program under the Grant Number 2011-1640 from the
EACEA of the European Commission, during which part of this work was
developed.}

\vspace{0.5cm}

%%%%%%%%%%%%%%%%%%%%%%%%%%%%%%%%%%%%%%%%%%%%%%%%%%%%%%%%%%%%%%%%
%%%%%%%%%%%%%%%%%%%%%%%%%%%%%%%%%%%%%%%%%%%%%%%%%%%%%%%%%%%%%%%%

%%%%%%%%%%%%%%%%%%%%%%%%%%%%%%%%%%%%%%%%%%%%%%%%%%%%%%%%%%%%%%%%

\end{document}